# An Overview of Load Balancing in HetNets: Old Myths and Open Problems

Jeffrey G. Andrews, Sarabjot Singh, Qiaoyang Ye, Xingqin Lin, and Harpreet Dhillon
Contact: jandrews@ece.utexas.edu

The University of Texas at Austin

July 15, 2013

**Abstract**

Matching the demand for resources ("load") with the supply of resources ("capacity") is a basic problem occurring across many fields of engineering, logistics, and economics, and has been considered extensively both in the Internet and in wireless networks. The ongoing evolution of cellular communication networks into dense, organic, and irregular heterogeneous networks ("HetNets") has elevated load-awareness to a central problem, and introduces many new subtleties. This paper explains how several long-standing assumptions about cellular networks need to be rethought in the context of a load-balanced HetNet: we highlight these as three deeply entrenched myths that we then dispel. We survey and compare the primary technical approaches to HetNet load balancing: (centralized) optimization, game theory, Markov decision processes, and the newly popular cell range expansion (a.k.a. "biasing"), and draw design lessons for OFDMA-based cellular systems. We also identify several open areas for future exploration.

## 1 Myth One: Signal Quality is the Main Driver of User Experience

Mobile networks are becoming increasingly complicated, with heterogeneity in many different design dimensions. For example, a typical smart phone can connect to the Internet via several different radio technologies, including 3G cellular (e.g. HSPA or EVDO), LTE, and several types of WiFi (e.g. 802.11g, n, or ac), with each of these utilizing several non-overlapping frequency bands. Cellular base stations (BSs) are also becoming increasingly diverse, with traditional macrocells often being shrunk to microcells, and further supplemented with picocells, distributed antennas, and femtocells. To the mobile user, who may be within range of many BSs or WiFi access points (APs)[1] over dozens of different frequency bands, all that really matters is whether some of them can jointly deliver the rate and latency that the user's applications require. Modeling and optimizing for this seemingly simple objective is in fact very challenging, and changes many entrenched ideas about wireless communication systems. We start with:

*Myth 1: The received signal-to-interference-plus-noise ratio (SINR) is the first-order predictor of the user experience, or at least of the link reliability. For example, the bit error rate follows a $Q(\sqrt{SINR})$ relation and data rate tracks $B \log(1 + SINR)$.*

This myth is deeply entrenched in the fields of communication and information theory, and indeed, even in the "five bars" display on virtually every mobile phone in existence. It was true conventionally, and still is "instantaneously". For example, the probability of correct detection for a given constellation is monotonically related to the detection-time SINR (i.e. any residual interference not removed by the receiver is treated as noise), as any communication theory text confirms. Outage is also usually thought of in terms of a target SINR, namely the probability of being below it. Further, information theory tells us that achievable data rate follows $B \log(1 + SNR)$, or $B \log(1 + SINR)$ if the interference is modeled as Gaussian noise, where $B$ is the bandwidth. Thus, increasing the data rate seems to come down to increasing SNR (or SINR) – which yields diminishing returns due to the log – or acquiring more bandwidth.

The critical missing piece is the load on the BS, which provides a view of resource allocation over time. Modern wireless systems dynamically allocate resources on the timescale of a millisecond, so even a 100 msec window (about the minimum perceptual time window of a human) provides considerable averaging. In contrast, classical communication and information theory as in the previous paragraph provide only a "snapshot" of rate and reliability. But the *user-perceived rate* is their instantaneous rate multiplied by the fraction of resources (time/frequency slots) they are allowed to use, which for a typical scheduling regime (e.g. proportional fair or round robin) is about $1/K$, where $K$ is the number of other active users on that BS in that band. This is pretty intuitive: everyone has experienced large drops in throughput due to congestion at peak times or in crowded events, irrespective of signal quality, e.g. "I have five

---

[1] Henceforth, we shall include WiFi APs as a type of BS: one using unlicensed spectrum and a contention-based

bars, why can't I send this text message?!" The technical challenge is that the load $K$ varies both spatially and temporally and is thus impossible to determine *a priori* for a particular base station. It is often hard even to find a good model for the load $K$: it is clearly related to coverage area, as larger cells will typically have more active users, but also depends on other factors like the user distribution, traffic models, and other extrinsic factors. A main goal of this paper is to introduce some recent approaches to load-aware cellular network models, along with an appreciation for the limitations of load-blind models.

## 2   Myth 2: The "Spectrum Crunch"

It is a nearly universal article of faith that the amount of electromagnetic spectrum allocated to wireless broadband applications is woefully inadequate. Indeed, in 2012 the President's Council of Advisors on Science and Technology released the report *"Realizing the Full Potential of Government-Held Spectrum to Spur Economic Growth"* explaining in detail the reasons more broadband spectrum is urgently needed, mirroring many of the observations and recommendations of the FCC's 2010 National Broadband Plan[2]. This leads us to:

*Myth 2: There is a "spectrum crunch", and global spectrum regulators urgently need to release a lot more spectrum for wireless broadband in order to improve the user experience.*

This myth can be immediately dispelled with the following observation. Globally, mobile data traffic more than doubled in 2012 for the fifth year in row, and this trend is universally predicted to continue for at least several more years. We called for a corresponding 1000x increase in cellular capacity back in early 2011 [1], which has subsequently been adopted as the primary objective of 3GPP [2] and Qualcomm's "1000x Data Challenge". The amount of useful spectrum available for broadband communication is about 1 GHz (in the US, about 550 MHz for cellular, 430 MHz for WiFi). Yet, in the most optimistic scenario, the FCC is considering releasing 500 MHz of new spectrum by 2020, which is not even 2x what was available as of 2010, and thus

---

access protocol, but still in principle able to serve the mobile users in question.

[2] Although these both focus on spectrum policy in the United States, with very few exceptions the US FCC has led major new initiatives regarding global spectrum usage.

yields a shortfall of more than 500x. Although there are good arguments for releasing more spectrum for wireless broadband usage, solving the current capacity crunch is not one of them.

Rather, what we have is an infrastructure shortage, not a spectrum shortage. Nearly everyone agrees that small cells should be added at a rapid pace to ease network congestion, and that this will be the key element to moving towards 1000x. However, the small cells (micro, pico, femto) will be deployed opportunistically, irregularly, and in fixed locations, and have a certain amount of resources they can provide (i.e. spectrum and backhaul). In stark contrast, the devices they serve move around, and sporadically request extensive resources from the network, while at other times are dormant. Thus, the load offered to each base station varies dramatically over time and space. Thus, a small cell network will require much more proactive load balancing in order to make good use of the newly deployed infrastructure.

Of course, despite the above myths, many others in both industry and academia have recognized the importance of including load in the analysis of rate. <u>The unifying point is that the modeling and optimization of load should be elevated to have a similar status as the amount of spectrum or the SINR.</u> However, doing so in a technically rigorous manner is not straightforward.

## 3   Technical Approaches to Load Balancing

Outside of communication systems, load balancing has long been studied as an approach to balance the workload across various servers (in networks) and machines (in manufacturing) in order to optimize quantities like resource utilization, fairness, waiting/processing delays, or throughput. In emerging wireless networks, due to the disparate transmit powers and base station capabilities, even with a fairly uniform user distribution, "natural" user association metrics like SINR or RSSI can lead to a major load imbalance. As an example, the disparity between a max SINR and an optimal (sum log rate wise) association in a three tier HetNet is illustrated in Figure 1 (a) and (b). As seen in the plot, in (a) macro BSs serve most of the users even when some small BSs are sitting idle, whereas in (b) the load is considerably more balanced.

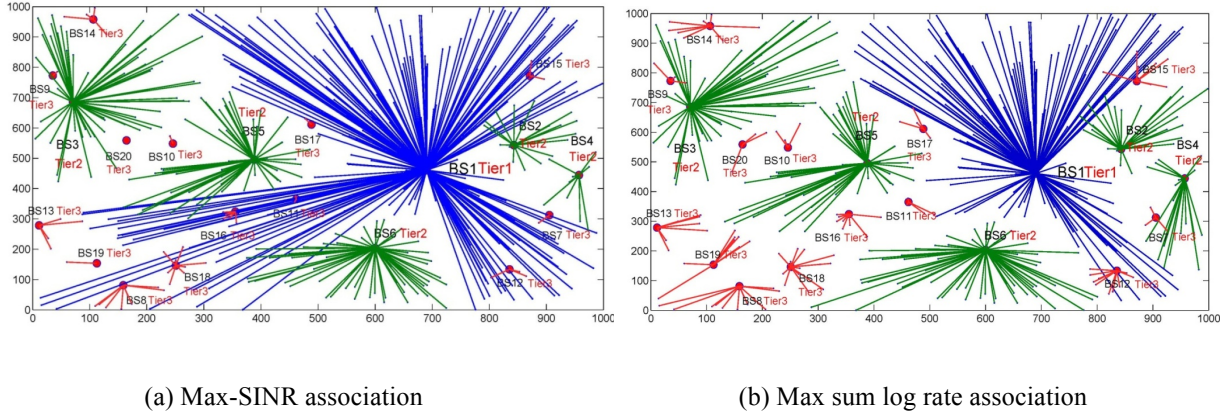

(a) Max-SINR association  (b) Max sum log rate association

Figure 1: Max-SINR association vs. max sum log rate association

Fundamentally, rate-optimized communication comes down to a large system-level optimization, where decisions like user scheduling and cell association are coupled due to the load and interference in the network. In general, finding the truly optimal user-server association is a combinatorial optimization problem and the complexity grows exponentially with the scale of the network, which is a dead end. We briefly overview a few key technical approaches for load balancing in HetNets.

## 3.1 Relaxed Optimization

Since a general utility maximization of (load-weighted) rate, subject to a resource or/and power constraint, results in a coupled relationship between the users' association and scheduling, this approach is NP hard and not computable even for modest-sized cellular networks. Dynamic traffic makes the problem even more challenging, leading to a long-standing problem that has been studied extensively in queuing theory, with only marginal progress made, known as the coupled queues problem.

One way to make the problem convex is by assuming a fully loaded model (i.e. all BS's always transmitting) and allowing users to associate with multiple BSs, which upper bounds the performance versus a binary association [3]. A basic form is to maximize the utility of load-weighted rate, subject to a resource or/and power constraint, where the binary association indicator is relaxed to a real number between 0 and 1. Following standard optimization tools, namely dual decomposition, a low-complexity distributed algorithm, which converges to a near-optimal solution, can then be developed. As it can be observed in Figure 2, there is a large (3.5x) rate gain for "cell-edge" users (bottom 5-10%) and a 2x rate gain for "median" users, compared to a maximum received power based association.

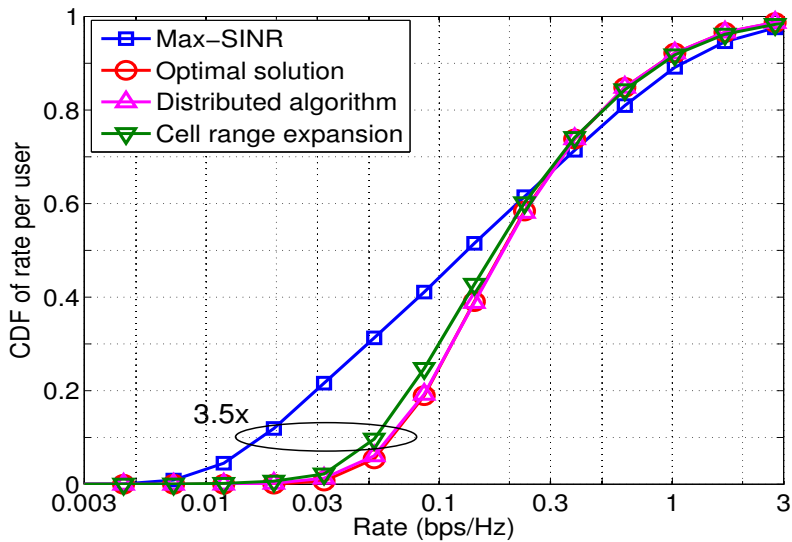

Figure 2: The distribution of rate using different association schemes. Cell range expansion is discussed in Section 3.4.

## 3.2 Markov Decision Processes

Markov Decision Processes (MDPs) provide a framework for studying the sequential optimization of discrete time stochastic systems in the presence of uncertainty. The objective is to perform actions in the current state to maximize the future expected reward. In the context of HetNets, MDPs have been used to study handoff between different radio access technologies (RATs), for example, cellular to WiFi offloading [4]. Another interesting application in HetNets is the association problem, e.g., in [5], a hybrid scheme where users are assisted in their decisions by broadcasted load information. However, as the size of the network increases, MDPs become harder to solve exactly. The MDPs also have limitations when dealing with continuous state spaces. An additional problem, in particular for complex, unstructured scenarios, is how to define adequate states and reasonable state transition model. Though in general, it is difficult to define an appropriate state model and solve it exactly for a large HetNet including different types of BSs as well as WiFi, taking the advantage of partly control from decision makers, MDP provides a possible approach for self-organizing HetNets to combine the benefits of both centralized and distributed design.

*3.3 Game Theory*

Game theory, as a discipline allows analysis of interactive decision-making processes, and provides tractable methods for the investigation of very large decentralized optimization problems. For example, a user-centric approach, without requiring any signaling overhead or coordination among different access networks, is analyzed in [6]. Another example is the study of dynamics of network selection in [7], where users in different service areas compete for bandwidth from different wireless networks. Although game theory is a useful tool, especially for applications in self-organizing/dynamic networks, the convergence of the resulting algorithms is, in general, not guaranteed. Even if the algorithms converge, they do not necessarily provide an optimal solution, which along with large overhead may lead to inefficient utilization. Further, since the main focus of game theory is on strategic decision-making, there is no closed-form expression to characterize the relationship between a performance metric and the network parameters. Thus, although we are not convinced that game theory is the best analysis or design tool for HetNet load balancing, it could provide some insight on how uncoordinated UEs and BSs should associate.

*3.4 Cell Range Expansion*

Biased received power based user association control is a popular suboptimum technique for proactively offloading users to lower power base stations and is part of 3GPP standardization efforts [8][9]. In this technique, users are offloaded to smaller cells using an association bias. Formally, if there are *K* candidate tiers available for a user to associate, then the index of the chosen tier is

$$k^* = \arg \max_{i=1...K} B_i P_{\text{rx},i} \qquad (1)$$

where $B_i$ is the bias for tier *i* and $P_{\text{rx},i}$ is the received power from tier *i*. By convention, tier 1 is the macrocell tier and has a bias of 1 (0 dB). For example a small cell bias of 10 dB means a UE would associate with the small cell up until its received power was more than 10 dB less than the macrocell BS. Biasing effectively expands the range/coverage area of small cells, so is referred to as *cell range expansion* (CRE).

A natural question concerns the optimality gap between CRE and the more theoretically grounded solutions previously discussed. It is somewhat surprising and reassuring that a simple per-tier biasing nearly achieves the optimal load-aware performance, if the bias values are chosen carefully [3] (see Fig. 2). However, in general, it is difficult to prescribe the optimal biases leveraging optimization techniques.

*3.5   Stochastic Geometry*

The previous tools and techniques seek to maximize a utility function $U$ for the *current network configuration*, for which we characterized the gain in average performance as

$$\mathbb{E}[\max_{\Omega} U] \qquad (2)$$

where $\Omega$ is the set of solution space. However, alternatively assuming an underlying distribution for the network configuration, another problem can be posed instead as in (3), where the optimization is over the averaged utility.

$$\max_{\Omega} \mathbb{E}[U] \qquad (3)$$

The latter formulation falls under the realm of *stochastic optimization,* i.e. the involved variables are random. The solution to (3) would certainly be suboptimal for (2) – and already we observed the gap between an optimized but static CRE and the globally optimal solution in the last section – but has the advantage of offering much lower complexity and overhead (both computational and messaging) versus re-optimizing the associations for each network realization.

Stochastic geometry as a branch of applied probability can be used for endowing base station and user locations in the network by a point process. By using Poisson point process (PPP) to model user and base station locations, in particular, tractable expressions can be obtained for key metrics like SINR and rate [11], which then can be used for optimization. This approach also has the benefit of giving insights on the impact of key system-level parameters like transmit powers, densities and bandwidths of different tiers on the design of load balancing algorithms. As an example of the applicability of this framework, cell range expansion has been analyzed using stochastic geometry in [12] by averaging over all the potential network configurations, revealing the effect of important network parameters in a concise form.

Modeling base stations as random locations in HetNets makes the precise association region and load distribution intractable. An analytical approximation for the association area was proposed in [12], which was then used for load distribution (assuming uniform user distribution) and consequently the rate distribution in terms of the per tier bias parameters can be found [12][13]. The derived rate distribution can then used to find the optimal biases simply by maximizing the biased rate distribution as a function of the bias value.

## 4   System Design Principles

We now explore several design questions that are introduced with load balancing. How much to bias? Can interference management help, how can it be done, and how much is the gain? As small cells will be continually rolled out over time, how (or does) the load balancing change as the small cell density increases? In this section we answer these questions, with the findings summarized in Table 1.

**Bias Values.** There are two major cases to consider for biasing: co-channel deployments (macro to small cell, in the same frequency band) and out-of-band biasing, such as cellular to WiFi. Both proactively push users onto BSs where they have weaker SINR, but there is a key difference. In the co-channel, not only is the received signal power decreased, but the interference is also increased, since it is by definition close to a strong source of interference (stronger than the new BS, else there would be no need to bias). In contrast, in out-of-band offloading, only the desired signal suffers, but in the new band the strong interference source is typically not present. Thus, optimal biasing is considerably more aggressive (e.g. 20 dB or more) in out-of-band offloading, as shown in Figure 3. In contrast, co-channel bias values are more like 5-10 dB, depending on the macro-pico transmit power differential.

**Blanking.** Following the logic in the previous paragraph, it seems that the optimal biasing values and resulting gains in co-channel deployments can be further increased if this co-channel macrocell interference could be avoided (in time or frequency) or cancelled. One such strategy is time-domain resource partitioning [9][10], where macro BSs are periodically muted. This is called *almost blank subframes* (ABS) in 3GPP LTE. Ideally, the offloaded users can then be scheduled in these blanked time slots, eliminating the co-channel macro tier interference. The operation of ABS in conjunction with range expansion is shown in Figure 4. Not surprisingly, when such a scheme is adopted, the biasing becomes much more aggressive, nearly in line with the out-of-band bias amounts. We can see in Fig. 5 that the optimal bias rises from about 6 dB up to 20 dB as the amount of blanking is increased, with an optimum around 16 dB for 5 picocells/macrocell. This assumes a scenario where the offloaded users are only served in the blanked time slots. Alternatively, if offloaded users can also be served in "normal" slots when the macros are on, then the optimal amount of blanking grows in proportion to the small cell density, as seen in Fig. 6. In either case, for plausible small cell deployments, the optimal amount of blanking is approximately one half. This strikes many as counter-intuitive but it is true: the macrocells (the apparent network bottleneck) should be shut off about half the time, because they are also the biggest interferers.

**Biasing as Small Cell Density Increases.** As small cells are increasingly dominant part of the cellular network, say in five years, will such aggressive biasing still be needed? The answer again depends on whether the offloading is co-channel or out-of-band. Increasing the small cell density increases the interference in both cases, but also the likelihood of being able to connect to a nearby small cell. In the out-of-band case, the increasing small cell interference makes connecting to a distant small cell less attractive, since the small cell interference is orthogonal to the macrocell. So in this case the optimal offloading bias decreases as the density increases. However, in the case of co-channel offloading, the small cell density does not affect the optimal offloading bias, because the interference they cause affects all users equally [13].

We conclude with our third myth (actually two combined into one), now dispelled by these results.

*Myth 3: Adding small cells at random requires sophisticated new interference management approaches so as not to undermine the carefully planned cellular network.*

Even randomly deployed BSs at arbitrary transmit power do not decrease SIR assuming a max-SIR association [11][12]. Since we have shown it is possible to do better than max-SIR, adding BSs can therefore only increase the rate CDF, even if the SIR is decreased (which it is, by definition, when departing from a max-SIR association). However, there is a grain of truth in this in the context of biasing.

*Reality: The benefit of interference management is increased with load balancing since the offloaded users now experience much larger interference than before.*

Because offloading does in general lower the SINR, there is the potential for an increased gain from interference management and cancellation. We observed one example in the blanking case, others could be conventional interference cancellation, or also from base station cooperation (CoMP).

**Table 1: Load Balancing Rules of Thumb**

|  | In-band offloading | In-band offloading with blanking | Out-of-band offloading |
|---|---|---|---|
| **Optimal small cell bias[3]** | 5-10 dB | 15-20 dB | 20-25 dB |
| **Increasing small cell to macrocell ratio** | Invariant | Optimal bias decreases, optimal fraction of blanked resources decreases. | Optimal bias decreases |

---

[3] The bias value given in this table is for a small cell density of five times that of macro cell and a transmit power difference of about 23 dB, and varies due to other modeling aspects such as propagation.

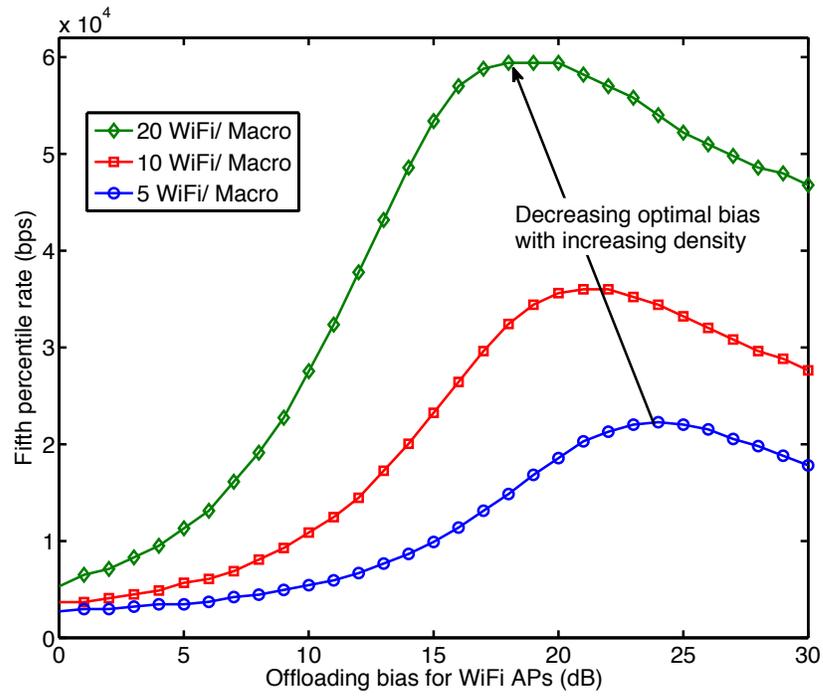

**Figure 3: Variation of fifth percentile rate with offloading bias for different WiFi AP densities (relative to the macrocell density)**

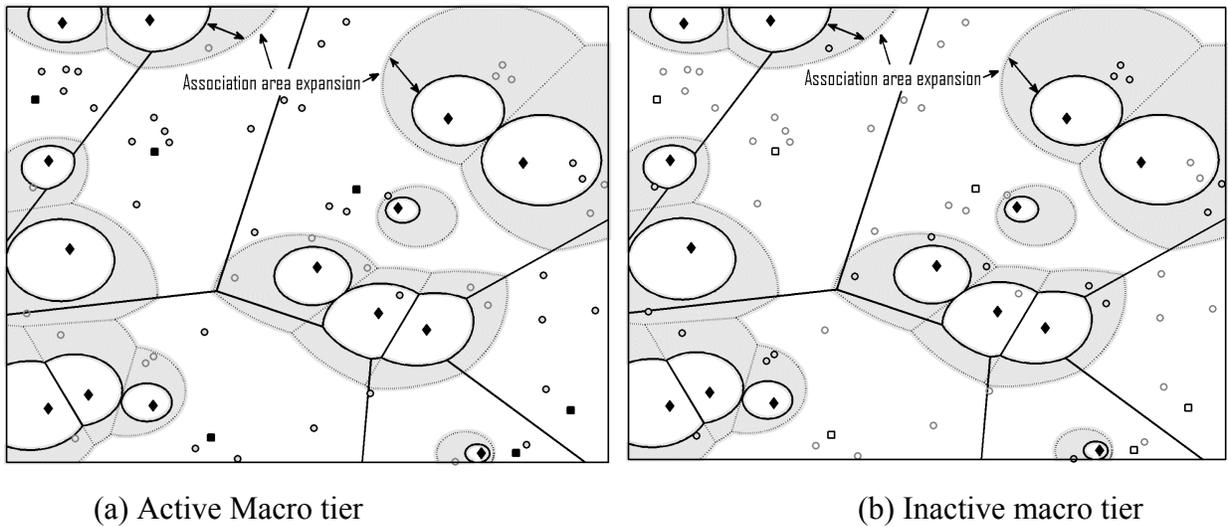

(a) Active Macro tier

(b) Inactive macro tier

**Figure 4: A filled marker is used for a node engaged in active transmission (BS) or reception (user). (a) The macro cells (filled squares) serve the macro users and small cells (filled diamonds) serve the non-range expanded users (filled circles). (b) The macro cells (hollow squares) are muted while the small cells (filled diamonds) serve the range expanded users (filled circles in the shaded region).**

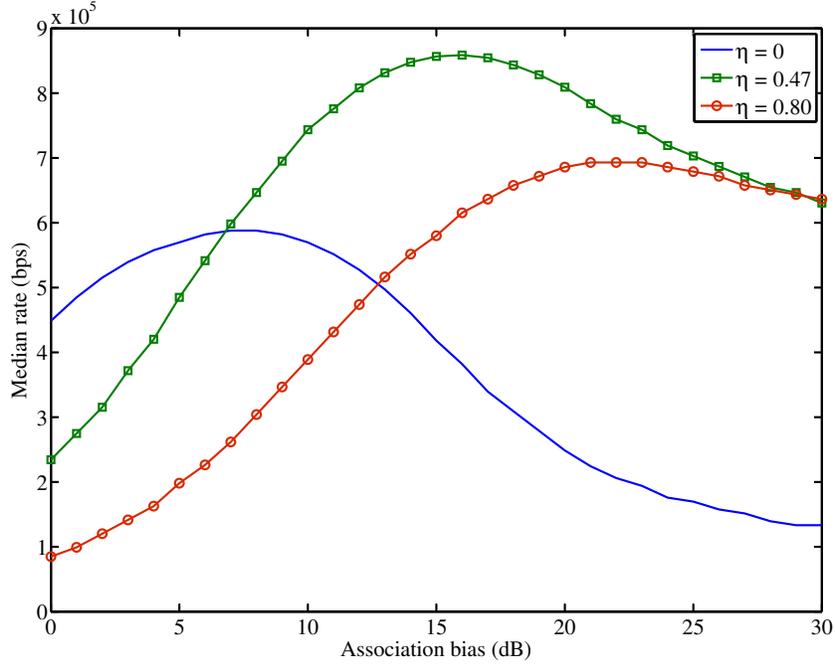

**Figure 5: Median user rate vs. bias with blanking ($\eta$ is fraction of blanked frames). 5 small cells per macro. Note how the optimal bias increases wit the amount of blanking.**

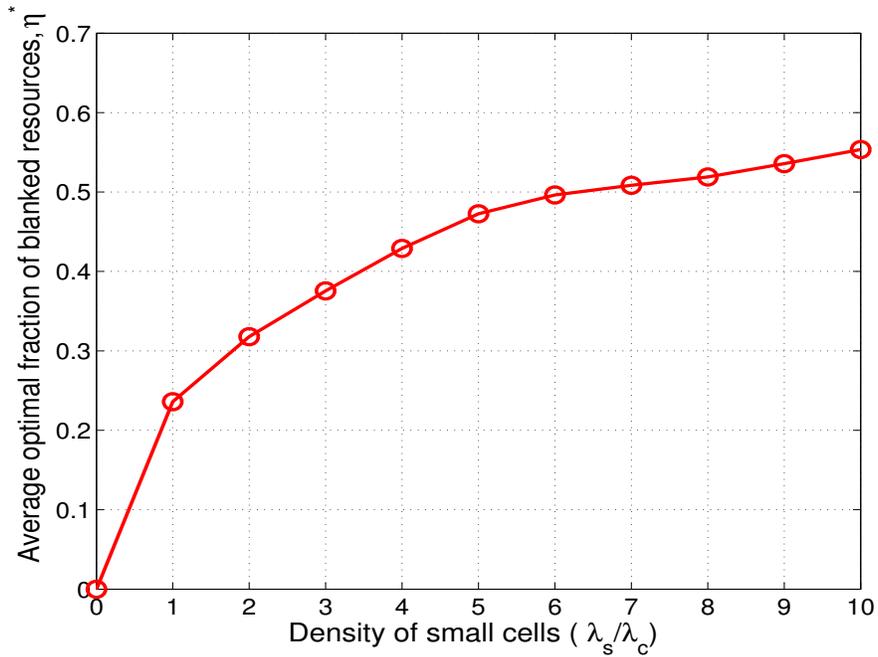

**Figure 6: Optimal blanking amount as small cell density increases. For a reasonable range, it appears macrocells should be shut off about half the time.**

# 5 Open Challenges

Load balancing for HetNets is far from being fully understood. What is clear is that it offers considerable flexibility and gain to the system designer, while calling into question several prior axioms for communication system theory. We conclude by offering some thoughts on fruitful avenues for future research and exploration.

## 5.1 Comprehensive Cell Range Expansion Study

Although the initial evidence appears very promising for cell range expansion to be a simple vehicle for realizing load balancing gains, there is still much work to do. To begin with, the analytical models used thus far often involve simplified assumptions, e.g. uniformly distributed UEs, omni-directional single-antenna transmission and reception, fixed transmit power, simple scheduling techniques, and so on. Some of these assumptions help make the analysis tractable, but may not be realistic. It would be useful to explore in depth the sensitivity of biasing and the ensuing gains to all these different aspects: some may be robust, others may not. For example, we saw above that out-of-band biasing should be an order of magnitude more aggressive than co-channel biasing.

In addition, we have been characterizing the network performance in an average sense, which allowed us to characterize per-tier "optimum" biasing. If it turned out that "optimum" biasing is quite sensitive to e.g. the spatio-temporal distribution of users, then a more sensible approach would be to adopt per-BS bias values, for example predicated on their current load.

## 5.2 Load Balancing with Implementation Constraints

Quite a few realistic factors/constraints of HetNets have been ignored in existing load balancing studies.

**The Backhaul bottleneck**: Small cells will often be backhaul-constrained; for example, the capacity of a femtocell or WiFi AP is usually limited by the wired backhaul connection. Taking this backhaul constraint into account, the amount of desired data offloading from macrocells may be reduced, particularly once the small cells are loaded beyond a threshold, which could be dependent on the backhaul [13]. A simple first approach would be to integrate the backhaul limitation into the associated bias value.

**Mobility**: Supporting seamless handovers among various types of cells in a HetNet is essential. In an ideal load balancing setting, a user of moderate or high mobility on entering a small cell association area should be offloaded from its original macrocell and back when it is no more near the small cell. However, it is known that handovers involve relatively complicated procedures as well as costly overhead. In the case of a short sojourn time in small cell, it may be preferable from a system-level view to temporarily tolerate a suboptimal BS association versus initiating a handover into and out of this cell. A related issue is open vs. closed access small cells.

**UE capability**: Despite its clear benefits, biasing UEs towards small cells does lower SINR. In LTE-A systems, the link throughput obtained under adaptive modulation and coding (AMC) with a typical codeset is zero when SINR is lower than about -6.5 dB. Thus, there are limits to offloading: a UE might theoretically get a better rate having a small cell's 10 MHz to itself with an SINR of -15 dB, but this is not viable if the UE cannot decode the lowest rate modulation and coding scheme that the BS can send. This further motivates interference management/ cancellation; but is a further constraint to consider when trying to accurately state the load balancing gain.

**Asymmetric downlink and uplink**: In the downlink, due to the large power disparities between BS types in a HetNet, macrocells have much larger coverage areas than small cells. In contrast, UEs can transmit at the same power level in the uplink regardless of the BS type. In addition, the downlink traffic is typically much heavier than the uplink traffic. In view of these asymmetries, the optimal downlink association need not be optimal for uplink transmission. Thus, it is necessary to extend existing downlink load balancing work to the corresponding uplink scenarios. Ideally, a joint load balancing study of the downlink and uplink should be performed.

*5.3 Interaction with Emerging Techniques such as Device-to-Device*

Since aggressive load balancing is somewhat of a new paradigm for cellular network design, new techniques need to be evaluated in this context. For example, 3GPP has recently initiated a study item on device-to-device (D2D) communication, which allows direct communication between cellular users, and thus can be viewed as an offloading technique. In a D2D-enabled HetNet, there exists D2D mode selection (i.e. whether a D2D link should be formed) in addition to user-AP association; this coupling significantly complicates the load balancing problem. How to jointly exploit small cell offloading and D2D offloading remains unknown.

*5.4 Regulatory Issues and Recommendations*

Considering the significant gains brought by HetNets, the regulatory focus should be on making it easier to deploy and use small cell infrastructure. This could include legal means to encourage (or force) municipalities or other landholders to allow picocell deployments with fair compensation; currently many want macrocell type rental fees for picocells which harms the business case. FCC actions could include freeing up less-coveted spectrum for wireless backhaul, coupling the auction of new spectrum to service providers with commitments to deploy more small cells, and strongly encouraging open access deployment for femtocells and WiFi (opening up WiFi alone would have a massive effect), perhaps through economic incentives. Although all these may sound daunting, but when compared with the politics of taking spectrum away from current incumbents in industry, the military, and other government agencies, perhaps seems more palatable.

From a technical point of view, as WiFi penetration increases, cellular (LTE-A) and WiFi networks should be able to handoff users seamlessly among them. The provisions in 3GPP like access network discovery and selection function (ANDSF) [14] for inter-RAT offload and smart AP selection in Hotspot 2.0 [15] are steps in the right direction. However, there is still a lot of room for improvement of the medium access control (MAC) layer efficiency in WiFi. We envision that WiFi will move over time towards a more cellular-like MAC with a backwards compatible OFDMA-based multiple access scheduler.

# 6 Acknowledgements

The authors gratefully acknowledge Huawei, Intel, Cisco, and Nokia Siemens Networks for their support of this work, and in particular to Amitava Ghosh (NSN) and Mazin Al-Shalash (Huawei) for their technical collaboration and many insights. This work also greatly benefitted from detailed discussions during visits with Qualcomm's HetNet group, Samsung's Dallas Technology Lab and Broadcom's WiFi and LTE groups.